\documentclass[onecolumn,aps,preprint,showpacs,amsmath,amssymb]{revtex4}
\usepackage{graphicx}
\usepackage{latexsym}
\usepackage{dcolumn} 
\usepackage{stmaryrd} 
 
\usepackage{bm}

\begin{document}
\newcommand{\eq}{\begin{equation}}                                                                         
\newcommand{\eqe}{\end{equation}}             

\title{Three dimensional compressible  Navier-Stokes equation - self-similar and traveling wave solutions} 

\author{ I.F. Barna}
\address{Wigner Research Center of the Hungarian Academy 
of Sciences , H-1525 Budapest, P.O. Box 49, Hungary} 

\date{\today}

\begin{abstract} 
We investigate the  three dimensional compressible Navier-Stokes and the continuity equations in Cartesian coordinates for Newtonian fluids. The polytropic equation of sate is used as closing condition. 
The key idea is the three-dimensional generalization of the well-known self-similar Ansatz which was already used for non-compressible viscous flow in our former study (Commun. in Theor. Phys. 56, (2011) 745). In the second method 
the three dimensional traveling wave Ansatz was applied.     
The geometrical interpretations of the trial functions are also discussed. 
 
\end{abstract}

\maketitle 
   
\section{Introduction}
To study the dynamics of viscous compressible fluids, the compressible Navier-Stokes (NS)  partial differential equation (PDE) 
together with the continuity equation have to be investigated.
In Eulerian description in Cartesian coordinates  these equations 
are the following:  
\begin{eqnarray}
\rho_t + div [\rho {\bf{v}}]  &=& 0, \nonumber \\ 
 \rho[{\bf{v}}_t  + ({\bf{v}}\nabla){\bf{v}}] &=& \nu_1 \triangle 
{\bf{v}} + \frac{\nu_2}{3}  grad\> div\> {\bf{v}} - \nabla p + a,      
\label{nav} 
\end{eqnarray}
where ${\bf{v}}, \rho, p, \nu_{1,2} $ and $ a $ denote respectively the three-dimensional velocity field, density, pressure, kinematic viscosities 
and an external force (like gravitation) of the investigated fluid.
To avoid further misunderstanding we use $a$ for external field instead of the letter  $g $ which is reserved for a self-similar solution. In the later we consider no external force,  so $a =0$.   
For physical completeness we need an equation of state (EOS) to close the equations. We start with 
the polytropic EOS $ p = \kappa \rho^n$, where $ \kappa$ is a constant of proportionality to fix the dimension and $n$ is a free real parameter. (n is usually less than 2)
In astrophysics, the Lane $-$Emden equation is a dimensionless form of the Poisson's equation for the gravitational potential of a Newtonian self-gravitating, spherically symmetric, polytropic fluid. It's solution is the polytropic EOS which we use in the following. The question of more complex EOSs will be discussed later on. 
Now $ \nu_{1,2}, a, \kappa, n $ are parameters of the flow. 
For a better transparency we use the coordinate notation ${\bf{v}}(x,y,z,t) = (u(x,y,z,t),v(x,y,z,t),w(x,y,z,t))$ and for the scalar  density variable $\rho(x,y,z,t)$ from now on. 
Having in mind the correct forms of the mentioned complicated vector operations, the PDE system reads the following: 
\begin{eqnarray}
\rho_t + \rho_xu  + \rho_y v + \rho_z w + \rho[u_x+ v_y+w_z]  = 0 \nonumber \\ 
\rho[u_t + uu_x + vu_y + wu_z] - \nu_1[u_{xx}+u_{yy}+u_{zz}] - \frac{\nu_2}{3}[u_{xx}+v_{xy}+w_{xz}]
+ \kappa n \rho^{n-1}\rho_x    = 0 \nonumber \\ 
\rho[ v_t + uv_x + vv_y + wv_z] - \nu_1[v_{xx}+v_{yy}+v_{zz}] - \frac{\nu_2}{3}[u_{xy}+v_{yy}+w_{yz}]
+ \kappa n \rho^{n-1}\rho_y  = 0   \nonumber \\ 
\rho[w_t + uw_x + vw_y + ww_z] - \nu_1[w_{xx}+w_{yy}+w_{zz}]  - \frac{\nu_2}{3}[u_{xz}+v_{yz}+w_{zz}]
+ \kappa n \rho^{n-1}\rho_z  =0.    
\label{nav2}
\end{eqnarray}
The subscripts mean partial derivations. Note, that the formula for EOS is already applied.
   
There is no final and clear-cut existence and uniqueness theorem  for the most general non-compressible NS equation. 
However, large number of studies deal with the question of existence and uniqueness theorem related to various viscous 
flow problems. Without completeness we mention two works which (together with the references) give a transparent
 overview about this field \cite{uniq,uniq2}. 

According to our best knowledge there are no analytic solutions   
for the most general three dimensional NS system even for non-compressible Newtonian fluids.
 
However, there are various examination techniques available in the literature with analytic 
solutions for the restricted problem in one or two dimensions.    
Manwai  \cite{manwai} studied the N-dimensional $(N \ge 1)$ radial 
Navier-Stokes equation with different kind of viscosity and pressure 
dependences and presented analytical blow up solutions.	
His works are still 1+1 dimensional (one spatial and one time dimension)
investigations.  Another well established and popular investigation method is based on 
Lie algebra. There are other numerous studies available.  
Some of them are for the three dimensional case,  
for more see \cite{lie}. Unfortunately, no explicit solutions are shown and 
analyzed there. Fushchich {\it{et al.}} \cite{fus} construct a complete set of ${\tilde{G}}(1,3)$-inequivalent Ans\"atze of codimension one for the NS system, they present 19 different analytical solutions for one or two space dimensions. 

 Recently, Hu {\it{et al.}} \cite{hu} presents a study where,  
symmetry reductions and exact solutions of the (2+1)-dimensional NS are calculated. 
Aristov and Polyanin \cite{arist} use various methods  like Crocco transformation,  generalized separation of variables or the method of functional separation of variables for the NS and present large number of new classes of exact solutions. 
Sedov in his classical work \cite{sedov}* derive analytic solutions for the tree dimensional spherical NS equation 
where all three velocity components and the pressure have only polar angle dependence ($\theta$). Even this kind of restricted symmetry leads to a non-linear coupled ordinary differential equation system with a very rich mathematical structure. Additional similarity reduction studies are available from various authors as well \cite{jiao,fakhar,angol}.
A full three dimensional Lie group analysis is available for the three dimensional Euler equation of gas dynamics, with polytropic EOS  \cite{nad} unfortunately without any kind of viscosity.
Analytical solutions of the Navier-Stokes equations for non-Newtonian fluid is presented  for one radial and one time dimension by \cite{kina}.  
 
In our study we apply two basic and physically relevant Ans\"atze, to find  a) self-similar and  
b) traveling wave solutions, to system (2). 

The form of the one-dimensional self-similar Ansatz is well-known \cite{sedov,barenb,zeld} 
\eq 
T(x,t)=t^{-\alpha}f\left(\frac{x}{t^\beta}\right):=t^{-\alpha}f(\eta), 
\label{self}
\eqe 
where $T(x,t)$ can be an arbitrary variable of a PDE and $t$ means time and $x$ means spatial 
dependence.
The similarity exponents $\alpha$ and $\beta$ are of primary physical importance since $\alpha$  represents the rate of decay of the magnitude $T(x,t)$, while $\beta$  is the rate of spread 
(or contraction if  $\beta<0$ ) of the space distribution for $t > 0 $.
The existence of a self-similar shape solution also means that the examined system has no internal time scale. 
The most powerful result of this Ansatz is the fundamental or 
Gaussian solution of the Fourier heat conduction equation (or for Fick's
diffusion equation) with $\alpha =\beta = 1/2$. These solutions are exhibited on Figure 1. for time-points $t_1<t_2$. 
There is a reasonable generalization of (\ref{self})  in the form of $T(x,t) = h(t)\cdot f[x/g(t)] $, where $h(t), g(t)$ 
are continuous functions. The choice of  $h(t)=g(t) = \sqrt{T-t}$  is called the blow-up solution, which 
means that the solution goes to infinity during a finite time duration. 
  
Leray \cite{leray} in his pioneering work in 1934 at the end of the manuscript put the question whether it is possible to construct 
self-similar solutions to the NS system in ${\bf{R}}^3$  in the form of $p(x,t) = \frac{1}{T-1}P(x/\sqrt{T-t})$ and 
${\bf{v}}(x,t) = \frac{1}{\sqrt{T-t}}{\bf{V}}(x/{\sqrt{T-t}})$. 
In 2001 Miller {\it{et al.}} \cite{miller} proved the nonexistence of singular pseudo-self-similar solutions
of the NS system in the above form. Okamato has given an exact backward finite-time blow-up self-similar 
solution via Leray's scheme.\cite{okoma}

Unfortunately, there is no any direct analytic calculation for the three dimensional self-similar generalization of this Ansatz in the literature. 
 
The applicability of Eq. (\ref{self}) is quite wide and comes up in various 
transport systems \cite{sedov,barenb,zeld,kers,barn,barna2}. 
\begin{figure} 
\scalebox{0.4}{
\rotatebox{0}{\includegraphics{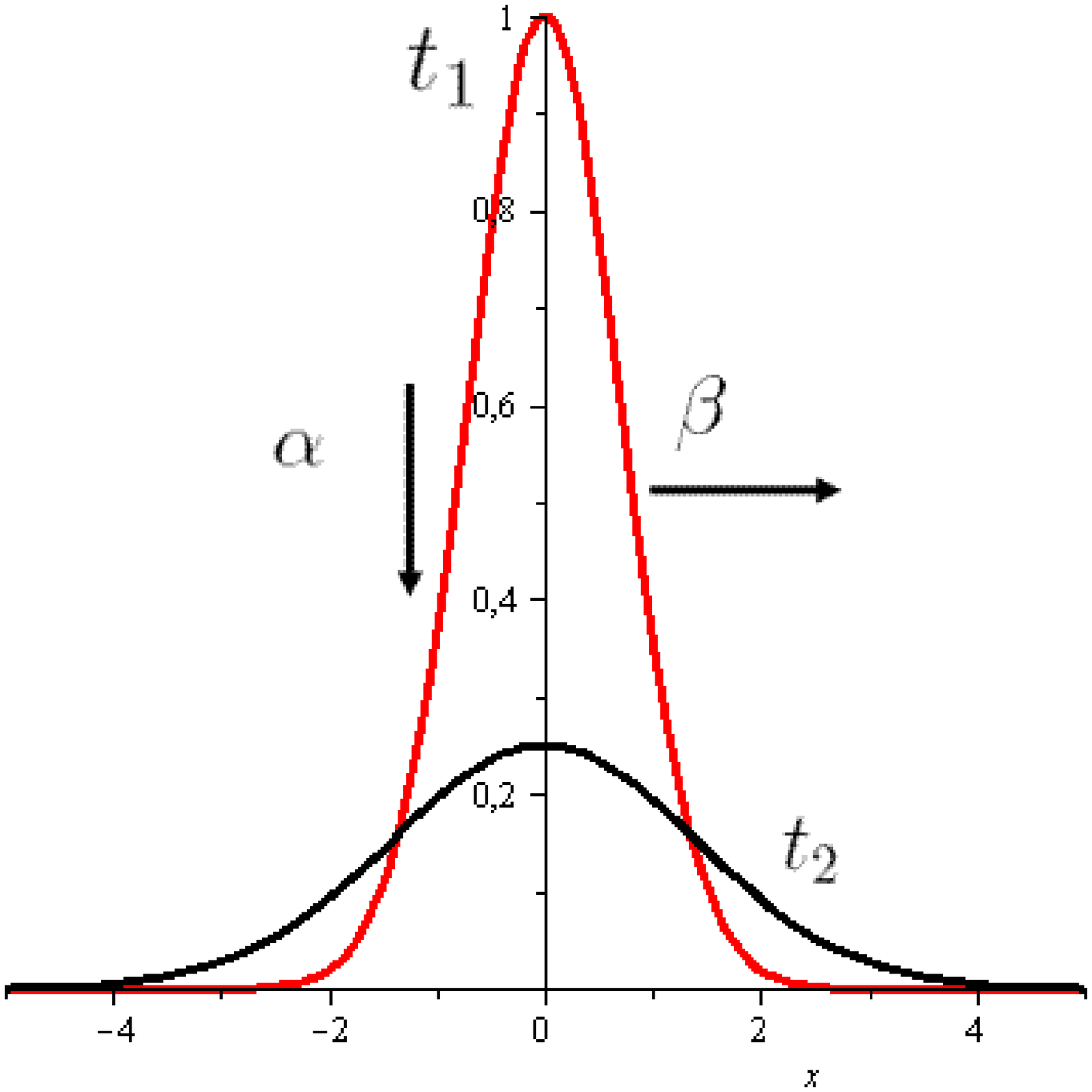}}}
\caption{A self-similar solution of Eq. (\ref{self}) for $t_1<t_2$.
The presented curves are Gaussians for regular heat conduction.}	
\label{egyes}       
\end{figure}
This Ansatz can be generalized for two or three dimensions in various ways.
One is the following 
\eq
u(x,y,z,t) = t^{-\delta}g\left(\frac{F(x,y,z)}{t^\beta}\right) = t^{-\delta}g(\eta),  
\eqe
where $F(x,y,z)$ can be understood as an implicit parameterization of a two dimensional surface. 
One of the most simple function is $F(x,y,z)=x+y+z=0$ which represents a plain passing through the origin. 
(The question of a more general plain, like $ax +by +dz +1 =0$, will be analyzed at the end of the manuscript.)
At this point we can give a geometrical interpretation of the Ansatz. Note that the dimension of $F(x,y,z)$ still has to be a spatial coordinate.  With this Ansatz we consider the velocity field $({\bf{v}_x} = u )$  - where the sum of the spatial coordinates lies on a plain - as a new entity. We do not consider all the ${\bf{R}}^3$ velocity fields but a plain of the ${\bf{v}}_x$ coordinate as an independent variable. This is the trick of the Ansatz. 
The NS equation  which is responsible for the dynamics maps this kind of velocities which are on this plain surface to another more complex geometry. 
In this sense we can investigate the dynamical properties of the NS equation in details.  
Our parametrization also means that there are as many realization of this Ansatz as many surfaces. Another reasonable and physically relevant surface would be   $ F(x,y,z) =  \sqrt{x^2+y^2+z^2}-r = 0$, which can be interpreted as an Euclidean vector norm or $L^2$ norm.  Now we contract all the $x$ coordinates of the velocity field $u$ (which are on a surface of a sphere with radius $r$)  to a simple spatial coordinate.  Unfortunately, if we consider the first and second spatial derivatives 
and plug them into the Navier-Stokes equation we cannot get a pure $\eta $ dependent ODE system some explicit $x,y,z$ or $t$ dependence tenaciously remain.  For a telegraph-type heat conduction equation, these Ans\"atze are useful to get solutions for the two dimensional case \cite{barna2}.  These investigations clearly show that for the 3d NS system  Eq. (2) only the plain surface is available. 

The second method which we apply  is the traveling wave Ansatz which gives a deeper insight into the wave properties of the investigated system such as the propagation speed (which can be even time dependent)
or the shape of the compact support of the wavefront.
More details about traveling waves can be found in the book of  
Gilding and Kersner  \cite{kers}, they have studied a large number of nonlinear diffusion-convection problems using this method.

\section{Self-similar solution}
Now, we concentrate on the first Ansatz Eq. (4) and search the solution of the Navier-Stokes PDE system in the following form:  
\begin{eqnarray}
\rho(x,y,z,t) = t^{-\alpha} f\left(\frac{x+y+z}{t^{\beta}}\right) =  t^{-\alpha} f(\eta),  \hspace*{3mm}
u(\eta) = t^{-\delta} g(\eta), 
\nonumber \\
v(\eta) = t^{-\epsilon} h(\eta), \hspace*{3mm} 
w(\eta) = t^{-\omega} i(\eta), 
\label{ans}
\end{eqnarray}
where all the exponents $\alpha,\beta,\delta, \epsilon, \omega $ are real numbers. (Solutions with integer exponents are called self-similar solutions of the first kind and can be obtained from dimensional argumentation as well.) 
According to Eq. (\ref{nav2}), we need to calculate all the first time derivatives of the velocity field, all the first and second spatial derivatives of the velocity field and the first spatial derivatives of the pressure. All these derivatives are not presented in details. More technical details of such a derivation is presented and explained in our former study \cite{imre2}.  
To get a final ODE system which depends only on the variable $\eta$, the following universality relations have to be hold  
\eq
\alpha = \beta = \frac{2}{n+1} \hspace*{1cm} \& \hspace*{1cm} \delta = \epsilon = \omega = 2 - \frac{4}{n+1}, 
\label{ns}
\eqe
where $n \ne -1$. 
Note, that the self-similarity exponents are not fixed values thanks to the existence of the polytropic EOS  exponent $ n$. 
(In other systems e.g. heat conduction or non-compressible NS system, all the exponents have a fixed value,
usually +1/2.) This means that our self-simlar Ansatz is valid for different kind of materials with different kind of EOS. 
Different exponents represent different materials with different physical properties which results different final ODEs with diverse mathematical properties.   

At this point we have to mention that even a more complicated EOS is possible, e.g. $ p \sim  f(\rho^l v^m)$. (The investigation of such problems will be performed in the near future but not in the recent study.)
  
Our goal is to analyze the asymptotic properties  of Eq. (\ref{ans}) with the help of Eq. (\ref{ns}).
 According to Eq. (3) the signs of the exponents 
automatically dictates the asymptotic behavior of the solution at infinite large time. 
All physical velocity components should decay at large times for a viscous fluid without external energy source term.  
The role of $\alpha $ and $\beta$ was explained after Eq. (3).  
 Figure 2 shows the $\alpha(n) $ and $\delta(n)$ functions. 
There are five different regimes:  
\begin{itemize}
\item
$n > 1 $ all exponents are positive - physically fully meaningful scenario -  spreading and decaying density and all speed components for large time - will be analyzed in details for general $n$    \vspace*{-4mm} 
\item
$n = 1$ spreading and decaying density in time and  spreading but non-decaying velocity field in time - not completely physical  but the simplest mathematical case  \vspace*{-4mm}
\item 
 $ -1 \le n \le 1 $   spreading and decaying density in time and  and spreading and enhancing velocity in time - not a physical scenario    \vspace*{-4mm} 
\item 
$n \ne -1$ not allowed case     \vspace*{-4mm}
\item 
$n \le -1  $ sharpening and enhancing density and sharpening and decaying velocity in time, we consider it an unphysical scene 
and neglect further analyzis.    
\end{itemize}

\begin{figure} 
\scalebox{0.42}{
\rotatebox{0}{\includegraphics{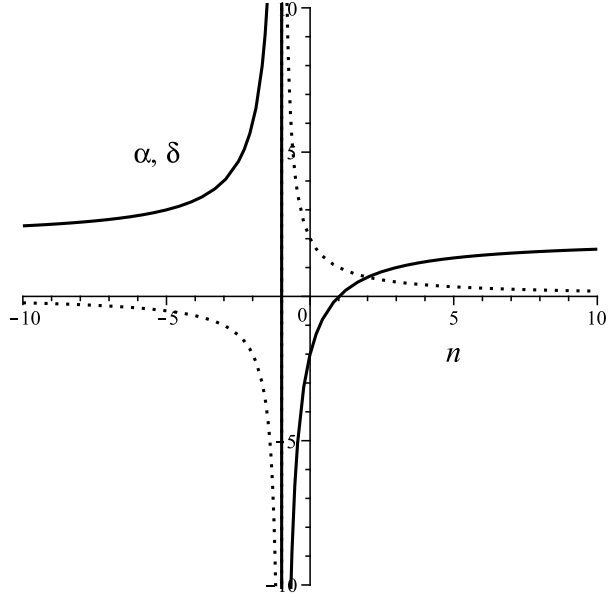}}}
\caption{Eq. (\ref{ns})  dotted line is  $\alpha(n) =2/(n+1)$ and solid line is $\delta(n)=2- 4/(n+1)$. }	
\label{kiteyo}       
\end{figure}




The corresponding coupled ODE system is:

\begin{eqnarray}
\alpha f + \beta f'\eta& = & f'[g+h+i] + f[g'+h'+i'] ,\nonumber \\ 
 f[-\delta g - \alpha \eta g' + gg' + hg' + ig']  &=&  -\kappa n  f^{n-1}f' + 3\nu_1g'' + \frac{\nu_2}{3}[g''+h''+i''] ,  \nonumber \\
 f[-\delta h- \alpha \eta h' + gh' + hh' + ih']  &=&  -\kappa n f^{n-1}f' + 3\nu_1h'' + \frac{\nu_2}{3}[g''+h''+i''] ,  \nonumber \\
 f[-\delta i -\alpha \eta i' + gi' + hi' + ii']       &=&  -\kappa n f^{n-1}f' + 3\nu_1i'' + \frac{\nu_2}{3}[g''+h''+i''] , 
\label{ode1}
\end{eqnarray}
where prime means derivation with respect to $\eta$. 
The first (continuity) equation it a total derivate (if $\beta = \alpha $) so we can integrate automatically getting 
\eq
\alpha f\eta =f [g+h+i] + c_0,    
\eqe
where $c_0 $ is proportional to the mass flow rate. Now, we simplify the NS equation with introducing only a single 
viscosity $ \nu = \nu_1= \nu_2.$ There are still too many free parameters remain for a general investigation, as we will see.   
Two various branches of solutions exist. 

The first one is for $ c_0 = 0$ which is the simpler case. 
Having in mind that the density of a fluid should be positive so $ f \ne 0$, 
we get $\alpha \eta = g+h+i$. 
With the help of the first and second derivatives of this formula Eq. (\ref{ode1}) can be reduced to the next non-linear
first order ODE 
\eq
-3\kappa n f ^{n-1} f' +  \left( \frac{2n-2}{n+1}\right) \left( \frac{2}{n+1} \right) \eta f = 0. 
\eqe
Note, that it is a first order equation, so there is a conserved quantity which should be a kind of general impulse in the parameter space $\eta$.  
We can also see that this equation has no contribution from the viscous terms with $\nu$ just from the 
pressure and from the convective term.   
The general solution reads  
\eq
f(\eta) = 3^{\frac{-1}{n-1}} \left(  \frac{ 2 \eta^2[n-1] }{\kappa n [n+1]} + 3c_1 \right)^{\frac{1}{n-1}}. 
\label{eredm}
\eqe 
Note, that for $ \{ n; n \in {\bf{Z}} \backslash \{- 1,0 \}\}$ exists  
$n$ different solutions  for $n>0$  (one    of them is the $f(\eta) = 0$)  and n-1 different solutions for  $n<0$ these 
are the n or (n-1)th roots of the expression. For  $\{n: n \in {\bf{R}}  \backslash \{-1,0 \}\} $ there is one real solution. 
(In the limiting case  $n=1$ (which means the $\delta = 0$) we get back the trivial result $ f = const $ which has no relevance.) 
For the $n =2$, the least radical case $f(\eta) =  \eta^2/(27 \kappa) +c_1$ which is a quadratic function in $\eta$ however,  the density function  $\rho = t^{-2/3}[ (x+y+z)^2/t^{4/3} ]   = (x+y+z)^2/t^2 $ has a proper time decay for large times.   

All the three velocity field components can be derived independently from  the last three Eqs. (\ref{ode1}). 
For the $v = t^{-\delta} g(\eta)$  the ODE reads:
\eq
-3\nu g'' + \left(2-\frac{4}{n+1} \right) gf - \kappa n f^{n-1}f' = 0. 
\eqe
Unfortunately, there is no solution for general $n$ in a closed form. However, for $n=2$ the solutions can be given by the WhittakerW and M functions \cite{abr}
(We fond additional closed solution only for $n=1/2$ and for $n=3/2$  which contains the HeunT functions, in a confusingly complicated expression.)
\eq
g =  \frac{\tilde{c}_1}{\sqrt{\eta}} \mbox{WhittakerM }\left( -\frac{\tilde{c}_1\sqrt{2\kappa}}{4\sqrt{\nu}}, \frac{1}{4}, \frac{\sqrt{2}\eta^2}{9\sqrt{\nu\kappa}}\right)  + \frac{\tilde{c}_2}{\sqrt{\eta}} \mbox{WhittakerW }\left(-\frac{\tilde{c}_2\sqrt{2\kappa}}{4\sqrt{\nu}}, \frac{1}{4}, \frac{\sqrt{2}\eta^2}{9\sqrt{\nu\kappa}} \right) + \frac{2}{3}\eta,
\label{wit}
\eqe
where $\tilde{c}_1$ and $\tilde{c}_2$ are integration constants.   
The Whittaker M is the irregular solutions and the Whittaker W functions is the regular one.  
The Whitaker funtions can be expressed via the Kummer's confluent hypergeometric functions see \cite{abr}  
 \begin{eqnarray}
\mbox{Whittaker} M(\lambda,\mu;z) = e^{-z/2}z^{\mu+1/2}M(\mu-\lambda+1/2,1+2\mu; z);  \nonumber \\ 
\mbox{Whittaker} W(\lambda,\mu;z) = e^{-z/2}z^{\mu+1/2}U(\mu-\lambda+1/2,1+2\mu; z). 
\end{eqnarray}
Is special cases when $\kappa = \nu/,2$ the Whittaker functions can formally be expressed using  other functions
(e.g. Bessel, Err)  when $\{c_1: c_1 \in {\bf{N}} \backslash \{ -2,-4\} \}$.  
Figure 3. presents the Whittaker M/W functions of (\ref{wit}) for the $c_1=1, \nu =0.2, \kappa = 4$ parameter set. 
It can be shown with the help of asymptotic forms that the velocity field  $u \sim t^{-1/3}[ M \> \mbox{or} \>  W(\cdot,\cdot;t^{-4/9})]$ decays for infinite time which is a physical property of a viscous fluid.  \\ 
At last at this point we mention that for the non-compressible three dimensional NS equation, the exponents are 
all $1/2$ except the decaying of the pressure field  which is 1. In that case the density field can be described with the help of the Kummer functions
\eq
f(\eta) = c_1 \cdot KummerU \left(-\frac{1}{4},\frac{1}{2},\frac{(\eta+c)^2}{6 \nu} \right)
+ c_2 \cdot KummerM \left( -\frac{1}{4},\frac{1}{2},\frac{(\eta+c)^2}{6 
\nu} \right) + \frac{c}{3} -\frac{2a}{3},
\eqe 
where $c_1$ and $c_2$ are the usual integration constants. 
This formula was analyzed in our last study\cite{imre2} in depth.

\begin{figure}  
\scalebox{0.45}{
\rotatebox{0}{\includegraphics{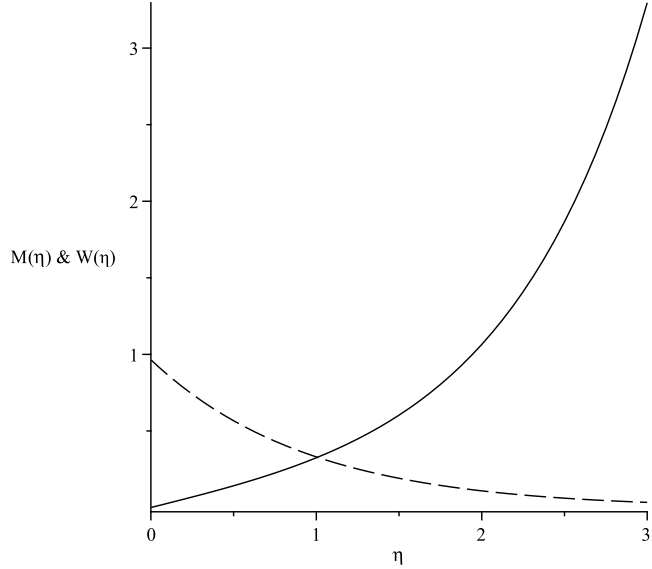}}}
\caption{The Whittaker M and W function of Eq. (\ref{wit}) for the $\tilde{c}_1=\tilde{c}_2=1, \nu =0.2, \kappa = 4$ parameter set. 
The solid line is the Whittaker M and the dashed line is the Whittaker W function.}
\end{figure}

The second case in our analysis is $c_0 \ne 0$.  
Now the following expressions can be derived from the continuity equation:  
 \eq 
g + h+i = \alpha \eta - \frac{c_0}{f}, \hspace*{1cm}
g' + h' +i' = \alpha + \frac{c_0 f'}{f^2}, \hspace*{1cm} 
g'' + h'' +i'' =    \frac{c_0 f''}{f^2} - \frac{2c_0f'^2}{f^3}, 
\eqe 
which helps us to obtain the final non-linear and non-autonomous second order ODE  
\eq
4 \nu c_0 f f '' - 8 \nu c_0 f'^2 + f' [ 3\kappa n f^{n+2} + c_0^2f ] + f^3 
\left[ c_0 \left( \frac{2}{n+1} \right) - c_0 f \left( \frac{2n-2}{n+1} \right) +   \left( \frac{4n-4}{(n+1)^2}\right)  \eta f  \right] = 0.
 \label{ODE3}
\eqe
Due to, the viscosity the first two terms are and the original pressure term gives contribution 
to the third one.   
This ODE contains four independent parameters. The viscosity $ \nu $, the polytropic exponent $n$, the strength of the pressure term $ \kappa $ and the $c_0$ integration constant from the continuity equation. From physical consideration all these parameters should be positive, except $c_0$.  
A complete analytical and numerical analysis of this equation is out of question. 
However, important properties can be examined. 
At the threshold  point where $n = 1 $( or $\delta = 0$) there exists an integrating factor $Q(\eta) = 1/f(\eta)^3$ which helps us to reduce the order of the ODE
\eq
f' + \frac{3\kappa}{4\nu c_0}f^3 + f^2 \left[ \frac{\eta}{4\nu} +c_1 \right] - \frac{c_0 f}{4 \nu} =0. 
\label{selfvegso}
\eqe
Unfortunately, even this ODE has no general analytic solution. Figure 4 presents the direction field for a typical 
$\kappa=c_0 = 1, \nu =0.01, c_1 = 0$  parameter set. 
Note, that there are basically two different trajectories available, a blow-up trajectory which goes to infinity at finite $\eta$ value and another one which has a maximum and a slow decay to zero. 
Larger $\nu$ value spreads out the steep step, however the $c_0  <0$ choice makes a reflection to the $f(\eta)$ axis.   
The $\kappa<0$ parameter value has no physical relevance. The lack of the density solution in a closed form automatically means the lack of he velocity field in a closed form as well. 


\begin{figure}  
\scalebox{0.45}{
\rotatebox{0}{\includegraphics{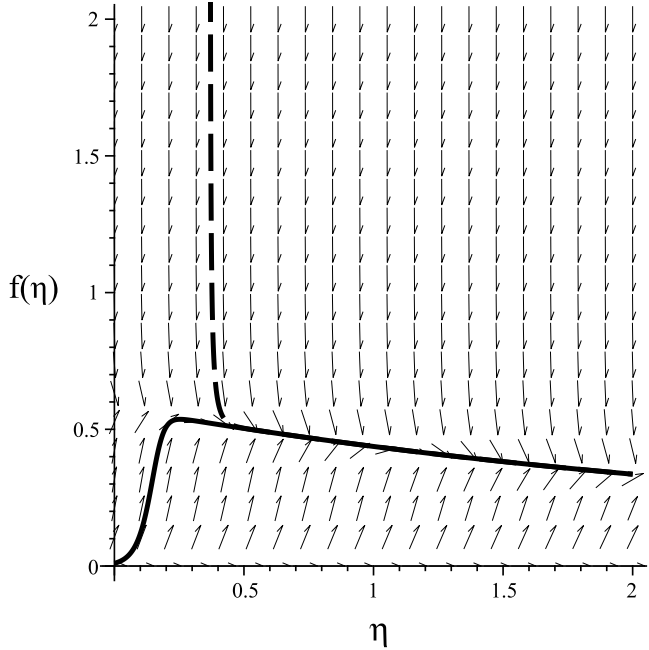}}}
\caption{The direction field of Eq. (\ref{selfvegso}) for $ \nu =0.01, \kappa =c_0=1,  c_1 = 0 $.  The dotted line presents the geometrical solution for the $f(0.4) =0.6$  and the solid line is for the $f(0)=0.01$ iniital conditions.}
\end{figure}

\section{Traveling wave solution}

Another physically relevant solutions can be found with the application of the  traveling wave Ansatz, 
which has the form of  $ \tilde{f}(\eta) = \tilde{f}(x \mp Ct)$ in one dimension. The minus sign means waves propagation to the right 
direction and the plus sign means the opposite. $C \ne0$ is a real number and means the propagation velocity. 
In our three dimensional study  we use the following generalization
\eq
\rho(x,y,z,t) = \tilde{f}(x+y+z-Ct) = \tilde{f}(\eta), \> u = \tilde{g}(\eta), \> v=\tilde{h}(\eta), \>  w=\tilde{i}(\eta). 
\label{hullam}
\eqe 
Similar notations are used for the shape functions as before. 
A geometrical interpretation is till available.  Now we consider all the particles which move perpendicular to the 
plain $x+y+z+Ct = 0 $ as a single entity. With this Ansatz not all the particles are considered in the entire
space  ${\bf{R}}^3$ but just from the plain above. 
  
 Applying this Anstaz to Eq. (2) we get an ODE system which is very similar to Eq. (7) 
\begin{eqnarray}
\tilde{f}'( \tilde{g}+ \tilde{h}+ \tilde{i}-C)& = &- \tilde{f}[ \tilde{g}'+ \tilde{h}'+ \tilde{i}']     \nonumber \\ 
 \tilde{f}[ \tilde{g}'( \tilde{g}+ \tilde{h}+ \tilde{i}- C)]  &=&  -\kappa n  \tilde{f}^{n-1} \tilde{f}' + 
 3\nu_1 \tilde{g}'' + \frac{\nu_2}{3}[ \tilde{g}''+ \tilde{h}''+ \tilde{i}'']   \nonumber   \\
 \tilde{f}[ \tilde{h}'( \tilde{g}+ \tilde{h}+ \tilde{i}-C)]  &=&  -\kappa n \tilde{f}^{n-1}\tilde{f}' + 
 3\nu_1 \tilde{h}'' + \frac{\nu_2}{3}[ \tilde{g}''+ \tilde{h}''+ \tilde{i}'']   \nonumber  \\
 \tilde{f}[ \tilde{i}'(\tilde{g}+\tilde{h}+\tilde{i}-C]       &=&  -\kappa n \tilde{f}^{n-1}\tilde{f}' + 
 3\nu_1 \tilde{i}'' + \frac{\nu_2}{3}[\tilde{g}''+\tilde{h}''+\tilde{i}''].    
\label{ode2}
\end{eqnarray}

The continuity equation can be integrated again yielding the following relation 
$\tilde{g}+ \tilde{h}+\tilde{i} = c - c_0/\tilde{f}$ where $c_0$ is an integration constant. 
There is a branching point again with respect to $c_0$. 

First case:  $c_0 = 0.$ After some trivial manipulations we get the final ODE of 
$ 3\kappa n \tilde{f}^n \tilde{f}' = 0 $
with the trivial solutions of $ \tilde{f} =0$ or $\tilde{f} = const.$
We can understand  these results as an application of a tricky coordinate system which moves together with the particles 
and therefore the density field  remains constant in this $\eta$ variable. 

The second case:   $c_0 \ne 0$. \\ 
Inserting the $ -c_0/\tilde{f} = \tilde{g}+\tilde{h}+\tilde{i}-C$ into the left hand side all the three NS equations can be integrated once getting and extra constant for each equation $c_1,c_2,c_3$. 
Adding all components together with some algebraic manipulation the last ODE can be derived   
\eq
-\frac{4}{3}\nu c_0 \tilde{f}' + 3\kappa \tilde{f}^{n+1} + \tilde{f}^2[c_4 -c_0 C] + c_0 \tilde{f}=0,
\label{final_wave}    
\eqe
where $c_4 = c_1 + c_2 +c_3$.
There are six free parameters in this equation,  $c_0,c_4, C$ the exponent $n$ the strength of the pressure term $\kappa$ 
and the viscosity $\nu$. The most important parameter is the exponent $n$ of the EOS. Dividing  (\ref{final_wave})
with $\nu c_0$ we can clearly see that the parameters (without $n$) only responsible for the strength of the last three terms.
With the choice of $C = c_4/c_0$ the third term cancels and a general closed form solution can be given
\eq
 \tilde{f}(\eta) =  \left( \tilde{c} e^{\frac{-3 n \eta}{4 \nu}}  - 3\kappa/c_0 \right)^{-1/n}
\label{solu}
 \eqe
where $\tilde{c}$ is a constant from this last integration.  
For physically reasonable positive parameters $ \tilde{c},\kappa, c_0, n > 0$, the function $\tilde{f}(\eta) $ has an upper range at $\eta = -\frac{4\nu}{3n} ln(3\kappa / c_0 \tilde{c}) $
however, for $n <0$, the function has no upper range and limit. 

The velocity field can be obtained from Eq. (\ref{ode2}). 
Inserting  $ -c_0/\tilde{f} = \tilde{g}+\tilde{h}+\tilde{i}-C$ into the left hand side of Eq. (\ref{ode2}) after integration we get 
 a linear-first order ODE 
\eq
3\nu \tilde{g}' +c_0 \tilde{g} - \kappa \tilde{f}^n + \frac{\nu}{3}\left(\frac{c_0}{ \tilde{f}^2} \tilde{f}' \right) + \tilde{c} =0, 
\eqe
 which can be solved by quadrature formally
\eq 
\tilde{g} = \left(\int  \frac{e^{\frac{c_0 \eta}{3\nu}} (-3 \kappa \tilde{f}^n  +\nu\frac{c_0 \tilde{f}'}{\tilde{f}^2} +3\tilde{c}  )}{-9\nu}  d\eta   + \tilde{C} \right)e^{\frac{-c_0 \eta}{ 3 \nu}} .
\eqe
Unfortunately,  the form of Eq. (\ref{solu}) does not make it possible to obtain a final closed form for the velocity field. 

At last, instead of Eq. (\ref{hullam})  the most general  plain is considered with the form   
of $\rho = f(ax+by+dz +1 -Ct)$ where $a,b,d$ are real numbers. 
The last term  in the last three  Eqs. (\ref{ode2})  contain $g''+h''+i''$ which rises a crucial problem (the terms with 
$a^a,b^2, d^2$ cannot be canceled), no simple ODE can be derived  for the density $f(\eta)$. Only a coupled nonlinear ODE system can be obtained 
  \begin{eqnarray}  
\tilde{g}' &=& \cdot \tilde{g} + \cdot \tilde{h}+ \cdot \tilde{i} +  \cdot 1/(ag+bh+di -C)^n \nonumber \\  
h' &=& \cdot g + \cdot h+ \cdot i +  \cdot 1/(ag+bh+di -C)^n \nonumber  \nonumber \\ 
i' &=&  \cdot g + \cdot h+ \cdot i +  \cdot 1/(ag+bh+di -C)^n 
\label{final3}
\end{eqnarray}
where $\cdot $ means twelve completely different constants built up from $a,b,d,C,\nu$ in a non trivial way ($n$ is still the exponent of the politropic EOS). We do not present the complete form of (\ref{final3}) because it gives no further information at this point.  It is clear that this ODE system has a strong singularity in the origin for $n>0$.  
(The investigation of this equation is out of our recent work.) With this last paragraph we just wanted to emphasize the 
complexity of the problem. Even a linear generalization of a linear function, introduction some parameters into the equation of the plain has a far reaching consequence which truly mimics the deeply non-linear feature of the NS equations. 

\section{Summary} 
In our study we investigated the compressible three dimensional Navier-Stokes equation with the self-similar and 
traveling wave Ansatz.  The existence of the polytropic EOS for the compressibility makes the calculations more complicated than for the non-compressible case. There is no general closed form of the self-similar solution available for the density and the velocity fields for any kind of material.  There are different scenarios available, some materials $n<1$ dictate unphysical exploding solutions, for $n>1$ both density and velocity fields show a decay property for large times which is the only reasonable solutions for dissipative systems.  There is a special case for $n=2$, where both fields can be expressed 
with the help of closed formulas which is our major result. The application of the traveling wave Ansatz gave us similar solutions as the self-similar Ansatz.   

We thank for Prof. Gabriella Bogn\'ar and Prof. Robert Kersner for useful discussions and comments. 




\begin{references}
 
\bibitem{uniq} J.G. Heywood, Acta Mathematica {\bf{136}}  (1976)  61. 

\bibitem{uniq2} R.B. Barrar,  Services Technical Information  Agency Comment Service Center AD 1711 Report, work done at Harvard University  under Contract N 5ori-07634,   1952. 

\bibitem{manwai} Y. Manwai, J. Math. Phys. {\bf{49}} (2008) 113102.   
	
\bibitem{lie} V.N. Grebenev, M. Oberlack and A.N. Grishkov,  
Journ. of Nonlin. Mathem. Phys. {\bf{15}} (2008) 227.

\bibitem{fus} W.I. Fushchich, W.M. Shtelen and S.L. Slavutsky, 
J. Phys. A: Math. Gen. {\bf{24}} (1990) 971.

\bibitem{grassi} V. Grassi, R.A. Leo, G. Soliani and P. Tempesta, 
Physica {\bf{286}} (2000) 79 *, ibid {\bf{293}} (2000) 421.  

\bibitem{hu}  X.R. Hu, Z.Z.  Dong, F. Huang  {\it{et al.}},
Z. Naturforschung A  {\bf{65}}  (2010)  504.  

\bibitem{arist} S.N. Aristov and A.D. Polyanin,  
Russ. J.  Math. Phys. {\bf{17}} (2010) 1.    

\bibitem{sedov} L. Sedov, {\it{Similarity and Dimensional Methods in Mechanics}} CRC Press 1993  
* (Page 120). 

\bibitem{jiao} J. Xia-Yu,  Commun. Theor. Phys. {\bf{52}} (2009) 389.

\bibitem{fakhar} K. Fakhar, T. Hayat, C.  Yi  and  T.  Amin,   Commun. Theor. Phys. {\bf{53}} (2010) 575.

\bibitem{angol}  D.K. Ludlow, P.A. Clarkson and A.P. Bassomx, J. Phys. A: Math. Gen. {\bf{31}} (1998) 7965–798.

\bibitem{nad} M. Nadjafikhah, http://arxiv.org/abs/0908.3598.

\bibitem{kina}   P. Cheng and T. Zhang, Appl. Math. J. Chinese Univ.  {\bf{24}} (2009) 483. 

\bibitem{barenb} G.I. Baraneblatt, {\it{Similarity, Self-Similarity, and 
Intermediate Asymptotics}} Consultants Bureau, New York 1979. 

\bibitem{zeld} Ya. B. Zel'dovich and Yu. P. Raizer {\it{Physics of Shock 
Waves and High Temperature Hydrodynamic Phenomena}} Academic Press, New York 
1966.

\bibitem{leray} J. Leray, Acta. Math. {\bf{63}} (1934) 193.  

\bibitem{miller} J.R. Miller, M. O'Leray and M. Schonbeck, Math. Ann. {\bf{319}} (2001) 809. 

\bibitem{okoma} H. Okamoto, Japan J. Indust. Appl. Math. {\bf{14}} (1997) 169. 

\bibitem{kers} B.H. Gilding and  R. Kersner, {\it{Traveling Waves in 
Nonlinear Diffusion-Convection Reactions,}} Progress 
in Nonlinear Differential Equations and Their Applications,  
Birkh\"auser Verlag, Basel-Boston-Berlin, 2004, ISBN 3-7643-7071-8. 

\bibitem{barn} I.F. Barna and R. Kersner, J. Phys. A: Math. Theor.
 {\bf{43}} (2010) 375210. 

\bibitem{barna2}  I.F. Barna and R. Kersner,  Adv. Studies Theor. Phys. {\bf{5}} (2011) 193.

\bibitem{imre2} I.F. Barna, Commun. in Theor. Phys. {\bf{56}} (2011) 745. 

\bibitem{abr} M. Abramowitz and I. Stegun, {\it{Handbook of Mathematical Functions}} Dover Publication., Inc. New York.                               
Chapter 13, Page 505. 

\end{references}
\end{document}